\documentstyle [12pt,epsf]{article}

\begin{document}
\baselineskip 7.5 mm

\def\thefootnote{\fnsymbol{footnote}}
\baselineskip 7.5 mm

\begin{flushright}
\begin{tabular}{l}
UPR-710-T \\
August, 1996 
\end{tabular}
\end{flushright}

\vspace{2mm}

\begin{center}

{\Large \bf 
The origin of pulsar velocities. 
}
\\ 
\vspace{8mm}

\setcounter{footnote}{0}

Alexander Kusenko\footnote{ email address:
sasha@langacker.hep.upenn.edu; address after October 1, 1996:
Theory Division, CERN, CH-1211 Geneva 23, Switzerland} 
and  
Gino Segr\`{e}\footnote{email address: segre@dept.physics.upenn.edu} 
\\
Department of Physics and Astronomy, University of Pennsylvania \\ 
Philadelphia, PA 19104-6396 \\

\vspace{12mm}

{\bf Abstract}
\end{center}

We have recently proposed an explanation for the birth velocities of
pulsars as resulting from asymmetries due to neutrino oscillations in the
cooling protoneutron star.  A specific prediction of this mechanism is that
the correlation of velocities and magnetic fields should be
stronger for slowly rotating neutron stars and weaker for those whose
periods are less than 1 s.  A remarkable agreement of this prediction
with the data favors the oscillation mechanism over the alternative 
explanations for the proper motion of pulsars.

\vfill

\pagestyle{empty}

\pagebreak

\pagestyle{plain}
\pagenumbering{arabic}
\renewcommand{\thefootnote}{\arabic{footnote}}
\setcounter{footnote}{0}

\pagestyle{plain}

Pulsars exhibit rapid proper motions \cite{ll,go} characterized by  
a mean birth velocity of $450 \pm 90$~km/s.  The origin
of these motions has been the subject of intensive studies and several
possible explanations have been proposed.  First, a substantial ``kick''
velocity may result from the asymmetries in the 
collapse, explosion, and the neutrino emission affected by convection
\cite{w}.  Second, evolution of close binary systems may also produce
rapidly moving pulsars \cite{b}.  Alternatively, 
it was argued \cite{ht} that the pulsar may be accelerated during the first
few months after the supernova explosion by its electromagnetic radiation.
The last two explanations seem to be disfavored by the data \cite{al,dc}. 

We have recently shown \cite{ks} that if one of the neutrinos has a mass of
a few hundred electron volts, the neutrino oscillations biased by the
magnetic field will cause a change in the shape of the neutrinosphere
making the momentum distribution of the outgoing neutrinos asymmetric.  
The resulting recoil velocity of the neutron star is in good quantitative
agreement with observations.  

If the ``kick'' velocities are, in fact, related to neutrino oscillations,
they provide a way to determine the neutrino mass in the range
currently inaccessible to collider and other  experiments. 
This would have important ramifications for both particle physics and
cosmology.  It is, therefore, important to focus on those aspects which may
distinguish this explanation from possible alternatives.  One specific
prediction of the oscillation mechanism is that the momentum asymmetry is
proportional to the magnetic field inside the neutron star.  As we show
below, for the pulsars with relatively large periods of rotation
(of the order of seconds), this
results in a correlation between the pulsar velocity $v$ and its magnetic
field $B$.  On the other hand, for those neutron stars which rotate
relatively fast, the $B$ -- $v$ correlation is obscured by the averaging of
the force acting on a pulsar and is much less significant statistically.
We demonstrate that 
this feature, that the $B$ -- $v$ correlation is strong for slowly rotating 
pulsars but is weak for faster ones, is in good agreement with the data.  
This is a strong evidence in favor of the oscillation mechanism \cite{ks}. 

Originally, the $B$ -- $v$ correlation \cite{ht,al} was reported for a
sample of 26 pulsars \cite{al} (irrespective of their periods of rotation)
and became the subject of several theoretical studies concerning its origin 
\cite{ht,al,expl_corr}.  However, as the information about new pulsars
became available, the $B$ -- $v$ correlation became only marginally
significant \cite{lla}.  (It was also observed \cite{lla} that there is no
correlation among the pulsars which are younger than 3~Myr.  We will 
reexamine the significance this observation below.)  
Another reason why this correlation may be important is because 
it can also help solve another long-standing puzzle, that of the
provenance of the $\gamma$-ray bursts \cite{ld}.  In this letter
we study the dependecy of the $B$ -- $v$ correlation on the rotation
speeds of pulsars and compare it to the theoretical predictions \cite{ks}.

The basic idea of the oscillation mechanism is the following.  
Neutrinos emitted during the cooling of a protoneutron star have  total
momentum, roughly, $10^2-10^3$ times the momentum of the proper motion of 
the pulsar.  A $0.1$\% to $1$\% anisotropy in the distribution of the
neutrino momentum would result in a ``kick'' velocity consistent with
observation.    In the dense neutron star an
electron neutrino, $\nu_e$, has a shorter mean free path than $\nu_\mu$,
$\nu_\tau$, or any of the antineutrinos.  If one of the latter, {\it
e.\,g.}, $\nu_\tau$, undergoes a resonant oscillation into $\nu_e$, above 
the $\tau$-neutrinosphere but below the $e$-neutrinosphere, it will be
absorbed by the medium.  Therefore, the {\it effective}
$\tau$-neutrinosphere in this case is determined by the point of resonance.    
The latter is affected by the magnetic field in such a way, that the
resonance occurs at different depth for the neutrinos emitted with a
velocity parallel to $\vec{B}$ as compared to those whose velocity is
anti-parallel to $\vec{B}$.  Therefore, the $\tau$-neutrinos will come out
from the regions of different temperatures, and will carry different
average momenta, depending on the direction of their velocity relative to 
$\vec{B}$.  

The resulting asymmetry in the momentum distribution is estimated to be 
\cite{ks} 

\begin{equation}
\frac{\Delta k}{k} = 0.001 \left ( 
\frac{3 \ {\rm MeV}}{T} \right )^2
\left ( \frac{B}{3 \times 10^{13} G} \right )
\label{res_ks}
\end{equation}

During the cooling stage of the protoneutron star, the $\tau$-neutrinos
come out with the average energy $\approx 10$ MeV \cite{snu_review}, which
corresponds to the temperature of $\approx3$ MeV.  We see that the observed
pulsar velocities, which require $\Delta k/k$ to lie between 0.001 and 0.01
can be explained by the values of $B$ from $3\times 10^{13}$ G to $3\times
10^{14}$ G. 

The magnetic fields at the surface of the neutron stars are estimated to be
of order $10^{12}-10^{13}$ G \cite{pulsar_review}.   However, a (toroidal)
magnetic field inside the pulsar may be as high as $10^{16}$ G 
\cite{pulsar_review}.  The existence of such a strong magnetic field is
suggested by the dynamics of formation of the neutron stars, by the 
stability of the poloidal magnetic field outside the pulsar, as well as by 
the fact that the only star whose surface field is well studied, the Sun,
has magnetic field below the surface which is $10^3$ larger than that
outside. (This strong magnetic field causes the sun spots when it
penetrates the surface.)  

It is clear from equation (\ref{res_ks}) that the deformations of the
neutrinosphere due to neutrino oscillations biased by the magnetic field 
can result in the asymmetry of the neutrino flux necessary to give the
pulsar a ``kick'' velocity consistent with the data.  

The direction of the recoil momentum is along the $\vec{B} $ direction and
is inclined to the axis of rotation at all times.  In the limit of a very
slowly rotating pulsar, the birth velocity should be correlated with the
strength of the magnetic field.  However, if the neutron star spins
rapidly about $\vec{\Omega}$, the angular velocity vector, 
the average force is non-zero only in the direction of $\vec{\Omega}$.  
If the angle between $\vec{B}$ and
$\vec{\Omega}$ is $\alpha(\vec{B},\vec{\Omega})$, the same  ``kick''
velocity will be generated for different values of $|\vec{B}|$, as long as
the product  $\left \{ |\vec{B}| \ cos \: \alpha(\vec{B},\vec{\Omega})
\right \}$ is the same.  Clearly, since $\alpha(\vec{B},\vec{\Omega})$ is
random, the $B$--$v$ correlation in this case can  be only as strong 
as the correlation between a random vector's length and its third component.  
Thus only the pulsars whose rotation is relatively slow will exhibit a
significant correlation between $B$ and $v$. 

The time scale is set by the duration of the neutrino emission and the rate
at which the neutrino flux diminishes.  The cooling of a protoneutron star
takes about ten seconds \cite{snu_review}, during which  the neutrino flux
continuously decreases.  
The characteristic time scale on which the neutrino flux
remains constant is, therefore, $T \sim $ few seconds.  If the period of
rotation $P \gg T$, one expects $B$ and $v$ to be correlated.  If, on the
other hand,  $P \ll T$,  the average force acting on a neutron star in
the direction orthogonal  to $\vec{\Omega}$ is zero, and the $B$ -- $v$
correlation is weak.

Unfortunately, there are not enough pulsars with large  periods to form a
statistically meaningful sample to test the correlation in the $P \gg T$
limit.  However, a generic prediction of the oscillation mechanism is that
the $B$ -- $v$ correlation should become increasingly more significant as
the period of rotation approaches a few seconds.  

\begin{figure}
\setlength{\epsfxsize}{4.5in}
\setlength{\epsfysize}{6.5in}
\centerline{\epsfbox{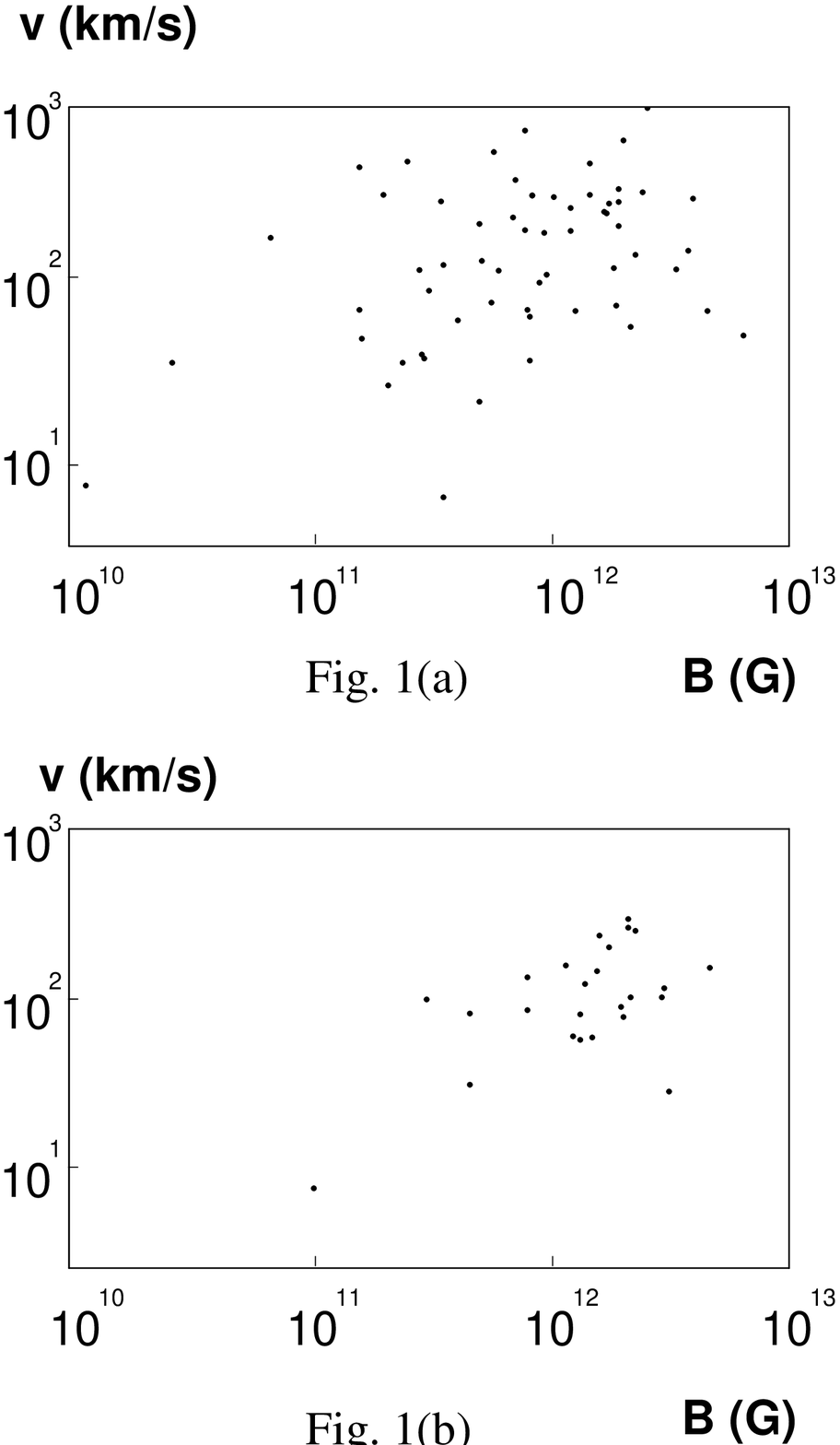}}
\caption{
The {$B$} -- {$v$} correlation is weak for the subset of 61 pulsars (a) 
with periods of rotation {$P < 1$} s, but is significant for the remaining 
26 pulsars (b), whose periods {$P > 1$} s.  
}
\label{fig1}
\end{figure}

This prediction is confirmed by the data as shown in Figure 1, where the 
data is plotted for a set of 87 pulsars \cite{hla} whose proper motions are
known.   For the pulsars with $P<1$~s (Fig. 1(a)), the correlator $c(B, v)
\equiv \langle B-\bar{B}, v-\bar{v} \rangle/ 
(\langle B-\bar{B}, B -\bar{B} \rangle \langle v-\bar{v}, v -\bar{v}
\rangle)^{1/2}=0.39$. On the other hand, for the pulsars with $P>1$~s,
$c(B,v)=0.58$ (Fig. 1 (b)), which is significantly greater. 
If one restricts the sample to those pulsars with $P>1.3$~s, the 
correlation increases further to $c(B,v)=0.77$.  The Spearman correlation
coefficients for the $P<1$~s, $P>1$~s, and $P>1.3$~s samples are 0.33, 0.44
and 0.58 respectively, all significant at more than $98 \%$ confidence 
level.  Such steady increase in the correlation with the period of rotation
is precisely what one would expect from the oscillation mechanism \cite{ks}.

The age $\tau_s$ of the pulsars may play an important role in the
statistical analysis because of various possible selection effects.  
It was argued \cite{lla} 
that the subset of pulsars younger than 3~Myr is a better representative of
the pulsar birth velocities.  We found similar increase in the $B$ -- $v$
correlation for this subset of pulsars: $c(B,v)=0.01$ for $P<1$~s, while 
$c(B,v)=0.59$ for $P>1$~s (although there are only 5 pulsars in the latter
category).  This is again in agreement with our theoretical predictions. 

We believe, however, that the set of pulsars with $\tau_s<3$~Myr may not
represent the distribution of birth velocities correctly. The reason is 
that for a pulsar born in the central bulge of the galaxy with an 
initial velocity of order 500~km/s, it will take about 3~Myr to travel 
the distance of $10^3$~pc characteristic of the size of the central bulge. 
Therefore, only  pulsars with high birth velocities would have emerged from
the galactic core (and from behind the bulge) during 
the first 3~Myr after birth.  The pulsars with lower birth velocities
would still be hidden behind the stellar matter in the core and their
signal would not reach the Earth.  If the birth velocities of the pulsars
are, in fact, correlated with their magnetic fields, one would expect only
the pulsars with large magnetic fields to be observable during the first 
3~Myr after birth.  This is consistent with the data presented by 
Lorimer, Lyne and Anderson in  figure 1(c) of their recent paper \cite{lla}
on the $B$ -- $v$ correlation.  Although, for the reasons just mentioned,
we do not consider the sample of younger pulsars a good representative for
the distribution of birth velocities, we emphasize that the effect in
question is evident for the subset of young pulsars, as well as for the
entire set. 

Finally, the increase in $B$ -- $v$ correlation with $P$ could be
attributed to some kind of a selection effect if the rotation periods of
the pulsars in our sample were correlated with their ages.  It is
reassuring, therefore, that there is no such correlation among the 87
pulsars we have considered.  

We have implicitly assumed that the magnetic field inside the neutron star,
between the electron and $\tau$-neutrinospheres, is proportional to the
magnetic field at the surface.  This is a plausible assumption
\cite{pulsar_review} but cannot be verified  experimentally. 
If the magnetic field outside the neutron star is not a good measure of the
fields strength inside, this could be the source of additional smearing of
the $B$ -- $v$ correlation. 

In summary, the substantial increase in the $B$ -- $v$ correlation 
for the slowly rotating pulsars lends further support to the recently
proposed explanation of the pulsar birth velocities based on neutrino
oscillations \cite{ks}.  The most significant correlation 
occurs for slowly rotating pulsars, while the overall $B$ -- $v$
correlation  is obscured by the inclusion of both slow and fast rotating
pulsars in the analysis.  

This work was supported by the U.~S.~Department of Energy Contract No.
DE-AC02-76-ERO-3071.

\end{document}